# Tailoring the energy landscape of a Bloch point singularity with curvature

Sandra Ruiz-Gomez[1*], Claas Abert[2,3], Pamela Morales-Fernández[1], Claudia Fernandez-Gonzalez[1], Sabri Koraltan[2,3,4], Lukas Danesi[2,3], Dieter Suess[2,3], Michael Foerster[5], Miguel Ángel Nino[5], Anna Mandziak[6], Dorota Wilgocka-Ślęzak[7], Pawel Nita[6,8], Markus Koenig[1], Sebastian Seifert[1], Aurelio Hierro Rodríguez[9], Amalio Fernández-Pacheco[10], Claire Donnelly[1,11*]

[1]Max Planck Institute for Chemical Physics of Solids, 01187, Dresden, Germany

[2]Faculty of Physics, University of Vienna, 1010 Vienna, Austria

[3]Research Platform MMM Mathematics-Magnetism-Materials, University of Vienna, 1010 Vienna, Austria

[4]ALBA Synchrotron Light Source, CELLS, Cerdanyola del Valles, 08290, Barcelona, Spain

[5]Vienna Doctoral School in Physics, University of Vienna, Vienna, Austria

[6]SOLARIS Synchrotron light Sources, 30-392 Crakow, Poland.

[7]Jerzy Haber Institute of Catalysis and Surface Chemistry, PAC, 30-239 Krakow, Poland

[8]Faculty of Physics, Astronomy and Applied Computer Science, Jagiellonian University, 30-348, Crakow, Poland.

[9]Depto. Física, Universidad de Oviedo, 33007 Oviedo, Spain

[10] Institute of Applied Physics, TU Wien, Wiedner Hauptstr. 8-10/134,1040 Vienna, Austria

[11]International Institute for Sustainability with Knotted Chiral Meta Matter (WPI-SKCM2), Hiroshima University, Hiroshima 739-8526, Japan

[*] Correspondence should be addressed to srgomez@ucm.es or claire.donnelly@cpfs.mpg.de.

**Abstract**

Topological defects, or singularities, play a key role in the statics and dynamics of complex systems. In magnetism, Bloch point singularities represent point defects that mediate the nucleation of textures such as skyrmions and hopfions. However, while the textures are typically stabilised in chiral magnets, the influence of chirality on the Bloch point singularities remains relatively unexplored. Here we harness advanced three-dimensional nanofabrication to explore the influence of chirality on Bloch point singularities by introducing curvature-induced symmetry breaking in a ferromagnetic nanowire. Combining X-ray magnetic microscopy with the application of in situ magnetic fields, we demonstrate that Bloch point singularity-containing domain walls are stabilised in straight regions of the sample, and determine that curvature can be used to tune the energy landscape of the Bloch points. Not only are we able to pattern pinning points but, by controlling the gradient of curvature, we define asymmetric potential wells to realise a robust Bloch point shift-register with non-reciprocal behaviour. These insights into the influence of symmetry and chirality on singularities offers a route to the controlled nucleation and propagation of topological textures, providing opportunities for logic and computing devices.

**Main Text**

Topology and topological textures, which mathematically describe fundamental properties of multidimensional objects, have made it possible to compare fields from gravitation [1-2] to liquid crystals [3-4] and magnetism [5-6], providing a simplified view to understand and compare the statics and dynamics of complex systems. One of the key reasons why topology has generated interest is due to the prospect of topological stability, where essential system characteristics persist through smooth transformations [7-8]. To facilitate the transformation of a system's topology, the introduction of defects, which disrupt the assumption of a continuous unit vector field, is required [9-12]. In three dimensions, these defects take the form of point defects that have been observed across diverse domains, including gravitational fields (black holes) [13], confined liquid crystals (commonly known as hedgehogs) [14-15] and magnetism (Bloch points) [16-17]. Despite their significant role in state transformations our understanding of these point defects remains limited due to the challenges in studying, isolating, and controlling them.

In this context, magnetism offers a unique opportunity to delve into the fundamental properties of such defects, allowing us to stabilise and image point defects known as Bloch point singularities [18-22] where, at length scales below 10 nm, the magnetisation vanishes [23]. Bloch points, which form as part of static magnetic configurations such as vortex rings, chiral skyrmion bobbers or Bloch point domain walls [24-26], play a key role in mediating in dynamic topological transformations such as the nucleation and evolution of skyrmions and the decay of hopfions in chiral magnets [27]. However, although the vast majority of studies of topological magnetism take place in chiral magnets, it is not yet understood how the chirality of a system could be used to stabilise, or indeed control the behaviour of these topological Bloch point defects.

To determine the influence of chirality on Bloch point singularities, we require two aspects: First, we need a precise control of the local symmetry that typically is achievable via the Dzyalozinskyi Moriya interaction (DMI) with interface-engineering, thin film or crystal growth. Second, we require the ability to reliably nucleate isolated Bloch points, which so far has only been possible in electrodeposited cylindrical nanowires, for which the materials are limited. However, until now combining both aspects has not been possible.

Here, we overcome this limitation by using geometric, i.e. curvature-induced symmetry breaking [28-30] and explore the stability and develop control of Bloch point singularities within a ferromagnetic system to gain insight on how local chirality affects their energy and behaviour. We modify the local chirality in a cylindrical nanowire by introducing curvature, allowing us to explore the effects of chirality on Bloch points.

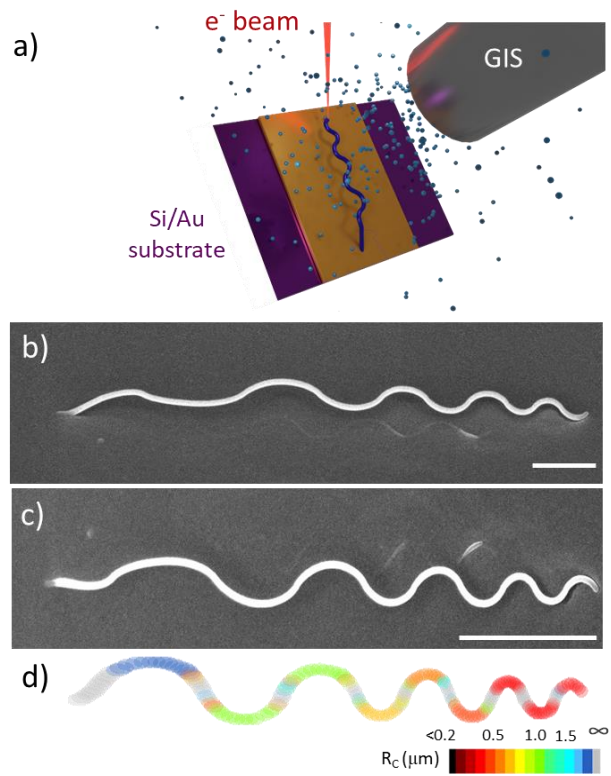

**Figure 1. 3D nanoprinting of complex 3D curved nanostructures**: a) Schematic of the FEBID process b) SEM image of an undulating structure fabricated with FEBID using the $Co_2(CO)_8$ precursor on a Si substrate coated with a Au film. The image has been taken with a tilt of 45° c) SEM image from a top view of the same structure and d) Schematics of the structures in which the color code indicates the radius of curvature along the structure. The scale bar is 1 μm in all figures.

We fabricated the curved cylindrical nanostructures with advanced 3D nanopatterning. Focused electron beam induced deposition combined with computer-aided design (f3ast software) makes possible the direct fabrication of intricate 3D structures with typical spatial resolutions on the order of a few tens of nanometers (see Figure 1a) [31-32]. Following this approach, we design and fabricate a nanowire with cylindrical cross section where we introduce regions with local radii of curvature spanning from 250 nm to 900 nm (Figure 1b-c), each distinctly colored in Figure 1d. These regions of curvature are separated by achiral straight segments, coloured in grey.

To determine the influence of the chirality on a Bloch point, we nucleate Bloch point domain walls, and then propagate them through regions of varying curvature, tracking their position. We track the position of the Bloch points using shadow X-ray photoemission electron microscopy (X-PEEM), where the shadow of the suspended nanowire is directly imaged in transmission [33]. We combine this imaging mode with the use of magnetic sample holders that allow the in-situ application of magnetic fields. A sketch of the measurement configuration is presented in Figure 2a. [34-35]

The three-dimensional nature of these structures, suspended above the substrate by a supportive leg (as depicted in the left part of Figure 1c), results in images with full visibility of their shadows. The shadows are elongated by approximately a factor of 3.5 along the beam direction due to the grazing incidence of the X-rays (16°). This provides an effective increase of the spatial resolution of the magnetic state in that direction, making it possible to resolve the structure of domain walls in suspended nanostructures. Figure 2b shows an X-ray absorption spectroscopy (XAS) image of an undulating structure measured at the Co $L_3$ edge, with the objective lens adjusted to ensure that the shadow of the structure remains in focus. Together with XAS images, dichroic images (XMCD images) can be measured by obtaining the pixel-by-pixel asymmetry between images measured with opposite X-ray helicities. As it can be seen in the sketch of Figure 2c, such an XMCD image gives a contrast proportional to the magnetization component along the X-ray direction, leading to an alternating XMCD contrast for an undulating structure in a single domain state.

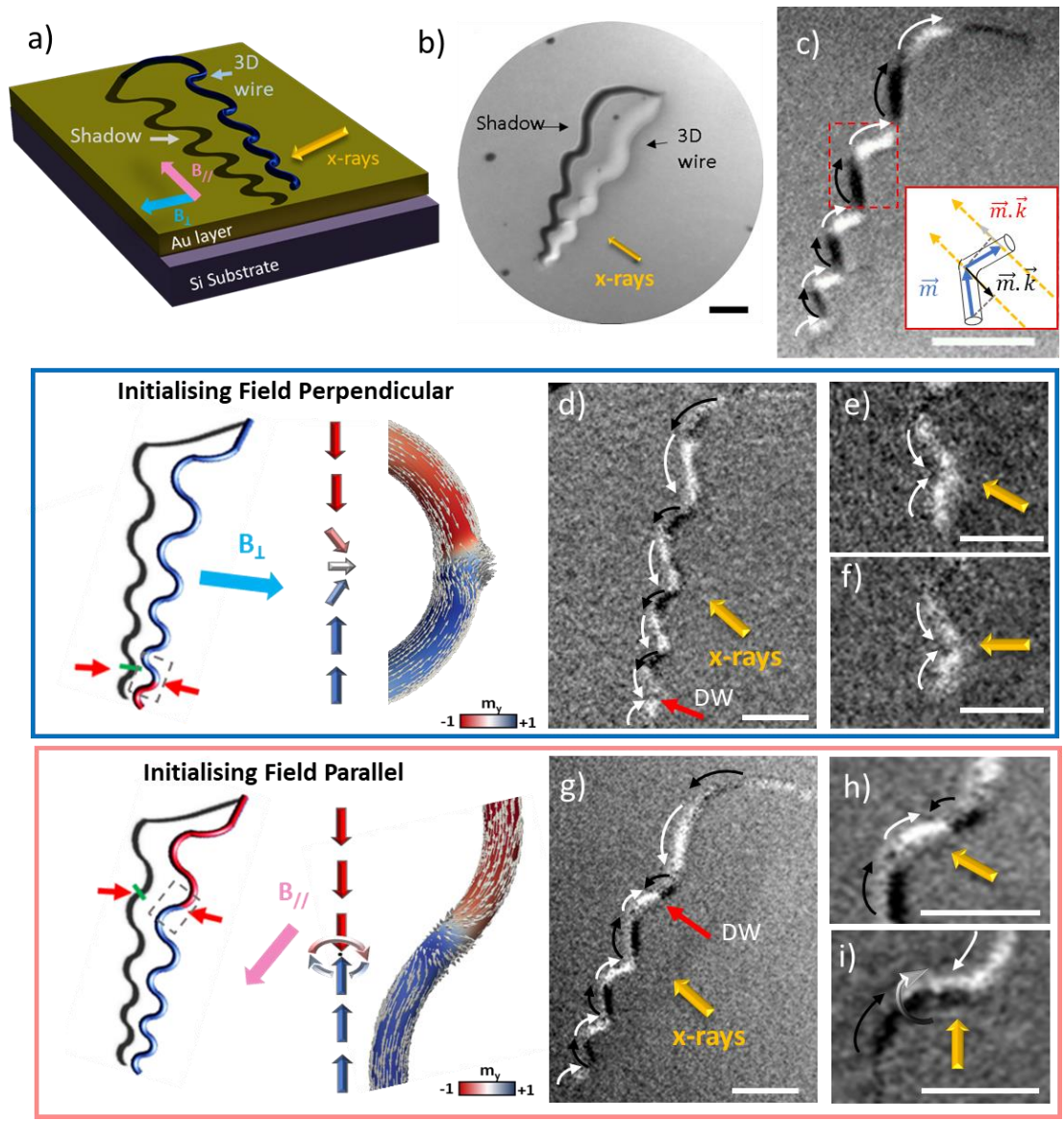

**Figure 2. Nucleation of TDW and BPDWs in curved nanostructures imaged by XMCD-PEEM**: a) Schematics of the measurement configuration of XMCD-PEEM in shadow mode. The nanostructure is illuminated with circularly polarized X-rays incident at an angle of 16° with the sample surface and the emitted photoelectrons are collected. The nanowire magnetic structure is analysed by photoelectrons emitted from the substrate by photons which have previously passed through the wire (transmission). b) XAS image of an undulating structure measured at the Co $L_3$ edge. c) XMCD image measured at the Co $L_3$ edge for the structure shown in b) in a single domain state with the magnetization pointing down along the structure. The sketch shows the projection of the magnetization with respect to the X-rays direction for the region marked with a red square. White (black) contrast in the XMCD images corresponds to a positive (negative) projection of the magnetisation onto the beam direction. d) XMCD image of the same structure after the nucleation of a domain wall applying magnetic fields perpendicular to the long axis of the nanostructure. The position of domain wall is denoted with a red arrow. e) Zoom of the previous image in the region where the domain wall is located. f) Zoom of the same domain wall after rotating the sample (25°) to measure with the X-rays oriented perpendicular to the wire. Yellow arrows indicate the direction of the incoming beam. g) XMCD image of the same structure after the nucleation of a domain wall by applying a magnetic field parallel to the long axis of the nanostructure. The position of the domain wall is denoted with a red arrow. h) Zoom of the previous image in the region where the domain wall is located. i) Zoom of the same domain wall after rotating the sample to measure with the X-rays oriented perpendicular to the wire. Yellow arrows indicate the direction of the incoming beam. The scale bar is 1 μm in all cases.

In order to induce the formation of a domain wall within the nanowire, two distinct methods can be employed: applying a magnetic field perpendicular to the long axis of the nanostructure (blue squares, Figure 2d-f), or parallel to it (pink squares, Figure 2g-i). In cylindrical nanowires, there are generally two types of domain walls that occur: transverse-vortex (TDW) and Bloch Point domain walls (BPDW) [12, 36-38]. TDWs are characterized by a rotation of the magnetization within the domain wall, perpendicular to the long axis of the nanowire [39]. In contrast, BPDWs exhibit a curling of magnetic moments within a plane perpendicular to the wire axis, featuring a Bloch point at their center [27]. So far, the effect of the local symmetry breaking induced by curvature has been considered only in the context of TDWs which are chiral objects. The local symmetry breaking can be thought of as introducing chirality into the system, and has been shown to select the chirality of TDWs [40-41].

In our experiments, to distinguish between the types of domain walls that can be generated, we measure XMCD images with X-rays directed perpendicular to the section of the nanowire housing the domain wall [27]. In the first scenario, where the domain walls are initialised by a saturating field perpendicular to the long axis, only a uniform white contrast is observed, indicating the presence of a TVDW with its magnetisation aligned with the curvature towards the right on the image (Figure 2e-f). This result agrees with previous studies that have shown that curvature affects the energy of transverse domain walls, breaking their degeneracy and favouring one rotational direction over the other. In the second case (when the magnetic field is applied parallel to the nanostructure), we observe a black-white contrast pattern, characteristic of the magnetization curling around the Bloch point (as shown in Figure 2i.). We note that, unlike the TDW, the BPDW is never found in the curved region but rather in the straight section between two curved regions. The BPDW remains in this position even under the influence of small magnetic fields of up to 20 mT (see supplementary information), indicating that it is a well-defined local minimum of energy.

To understand the influence of curvature-induced symmetry breaking on the energetics of the BPDW, we perform finite element micromagnetic simulations [42], determining the evolution of the energy of a BPDW as it moves through curved and straight regions of a nanowire conduit. We simulate a 70 nm diameter cylindrical nanowire with two regions of different magnitude of curvature, as shown in Figure 3a. BPDWs were initialised at different positions between the centre points of the two curved regions (Figure 3b), and the lowest energy path (LEP) between the two states was found by applying the string method in a full micromagnetic model [42,43].. The energy landscape is shown in Figure 3c, where one first observes that as the domain wall propagates from the region of lower curvature (left, I) to the region of higher curvature (right, III) via an effectively straight section (II), the energy of the domain wall drops sharply, indicating that there is an energy minimum corresponding to the BPDW sitting in the straight region of the structure (II). The underlying cause of this energy minimum becomes clear when we consider the curvature-dependence of the exchange and magnetostatic energies separately (shown in SI). The exchange energy of the domain wall in general decreases with increasing curvature, consistent with curvature-induced pinning of transverse walls, due to the reduction in the total rotated angle of the magnetisation across the domain wall. In contrast, the magnetostatic energy increases with increasing curvature and, most noticeably, exhibits a sharp drop when the domain wall is in the effectively straight interface region where the curvature is 0 ($\kappa=0$). Since the total energy is dominated by the magnetostatic energy, the straight interface region represents a local energy minimum, and thus a pinning point for the domain wall.

The curvature-evolution of the magnetostatic energy can be understood considering the local magnetostatic charges of the domain wall. Specifically, the BPDW forms in systems where the magnetostatic energy dominates, to minimise surface charges at the cost of local volume charges and exchange energy in the vicinity of the Bloch point singularity. In straight regions, the symmetry and shape of the domain wall remain unchanged. However, as the domain wall propagates through curved regions, the breaking of symmetry leads to the increase in magnetostatic energy. This increase in magnetostatic energy dominates the overall energy, i.e. the impact of curvature can be considered a magnetostatic effect. An intuitive picture of this effect can be obtained by considering the symmetry of the Bloch point domain wall: indeed, the structure is an achiral, symmetric magnetic texture, in contrast to the transverse domain wall which is intrinsically chiral. Just as a transverse domain wall will have a lower energy in a region of preferred curvature-induced chirality, the energy of the achiral Bloch point domain wall will be lower in the achiral, straight region of the structure.

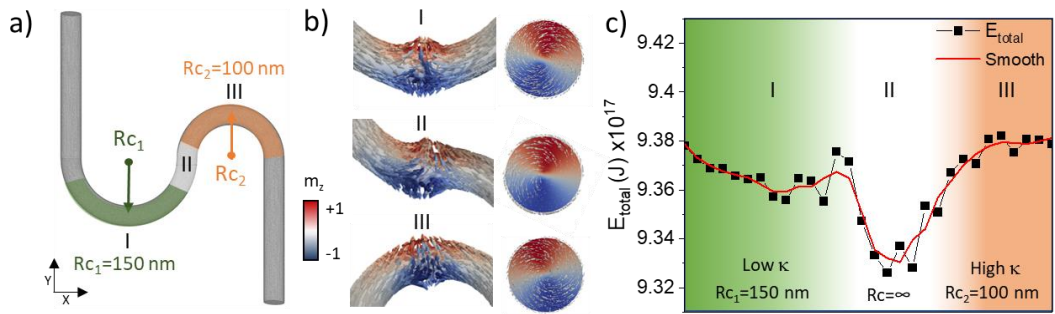

**Figure 3: Simulated energy landscape of a BPDW at curved and straight sections of a nanowire**. A schematic of the geometry is shown in a), with example positions of BPDWs and their central cross sections shown in b. c) The energy landscape reveals a well-defined potential minimum in the straight region (II, white), with the total energy of the BPDW increasing in regions of increasing curvature.

To experimentally confirm the curvature-dependence of the Bloch point domain wall energy, we propagate the domain wall through the nanostructure consisting of regions of alternating, increasing curvature with an external magnetic field. We apply 1 s pulses of magnetic field of increasing magnitude parallel to the long axis of the structure and track the position of the domain wall between pulses at remanence, as shown in Figure 4a. As we progressively increase the magnitude of the applied magnetic field pulses, the domain wall propagates along the structure in discrete steps, moving from one straight region to the next, as shown in Figure 4a. In fact, among a total of 29 observations of propagating domain walls, we consistently observed them pinned within the straight regions, corroborating our observation that the Bloch point exhibits a preference for symmetric, achiral regions (refer to supplementary information for further details and simulations).

As the BPDW propagates along the structure, the degree of curvature in the curved regions increases. This decrease in the radius of curvature is reflected in the strength of the depinning field, which also increases along the structure length. In particular, we observe a non-linear increase of the depinning field with curvature (Figure 4b). When we account for the structure's geometry [44], specifically the component of the field perpendicular to the domain wall, we observe that the depinning field increases linearly with the curvature (see Figure 4d), demonstrating a route to design the energy landscape of the Bloch point domain wall with curvature--induced symmetry breaking.

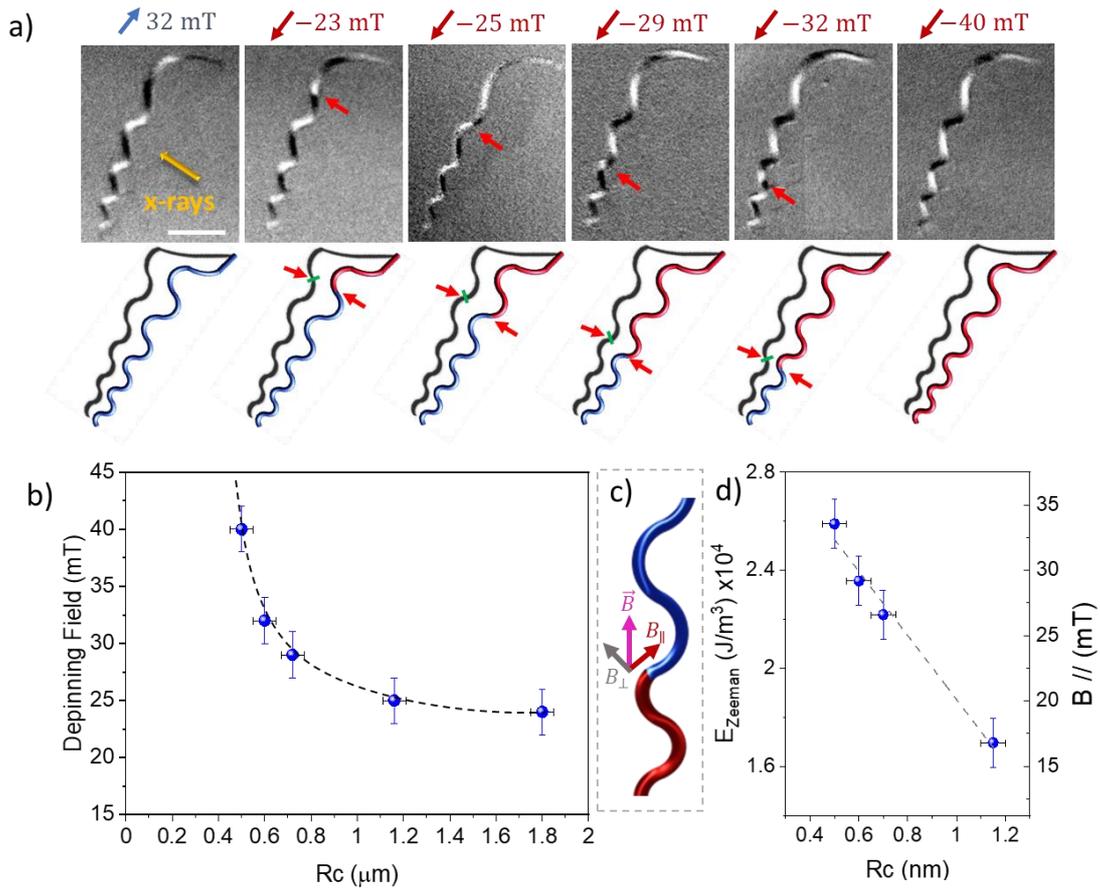

**Figure 4. Experimental demonstration of curvature induced pinning of a BPDW.** a) Magnetic configuration of the undulating structure after the application of magnetic fields. The upper row shows the XMCD images measured at the Co $L_3$ edge for the same structure before saturation (I) and after applying a magnetic field sequence of −24 mT (II), −25 mT (III) -29 mT (IV) , −32 mT (V) and -40 mT (VI). The XMCD signal is coming from the shadow of the structure where black and white contrast means a component of the magnetization parallel or antiparallel to the X-rays direction. The lower row shows schematics of the magnetization of the structure for each XMCD image where red indicates magnetization pointing up and blue corresponds to magnetization pointing down. The scale bar is 1 μm in all images. b) Depinning field as a function of radius of curvature extracted from a sequence of experiments as the one shown in panel a. c) Sketch of the component of the magnetic fields with respect to the domain wall. d) Component of the magnetic field parallel to the wire as a function of the radius of curvature and Zeeman energy calculated from the depinning field showed in panel b as a function of the radius of curvature.

So far, we have explored the unidirectional movement of domain walls, thus probing one side of the potential wells of the Bloch points. However, as the local curvature defines the energy barriers in our system, it is possible for the potential wells to exhibit asymmetry due to neighbouring energy barriers of different heights, as predicted by the micromagnetic simulations in Figure 3. This asymmetry would manifest as a non-reciprocity in the depinning fields of the domain wall, making it easier to propagate in one direction than in the other. We determine the asymmetry of the potential wells at the pinning points in the straight regions of the nanowire by measuring the depinning fields of the BPDW for both positive and negative field directions. Indeed, for a Bloch point in a straight region situated between two curved regions of varying magnitude (image 2 of Figure 5a), the propagation field required in one direction is higher than in the other, thereby exposing the inherent asymmetry within the potential well. Repeating this along the nanostructure, we map the energy barriers corresponding to individual curved regions, finding that they are symmetric and the height of the energy barrier can be set by varying the magnitude of the curvature. Consequently, the energy landscape of the nanostructure, as illustrated in Figure 5c, is composed of symmetric energy barriers interspersed with asymmetric wells, resulting in an inherently asymmetric propagation pattern of the domain wall I.e. a magnetic ratchet shift register, entirely determined by the structural geometry [45,46].

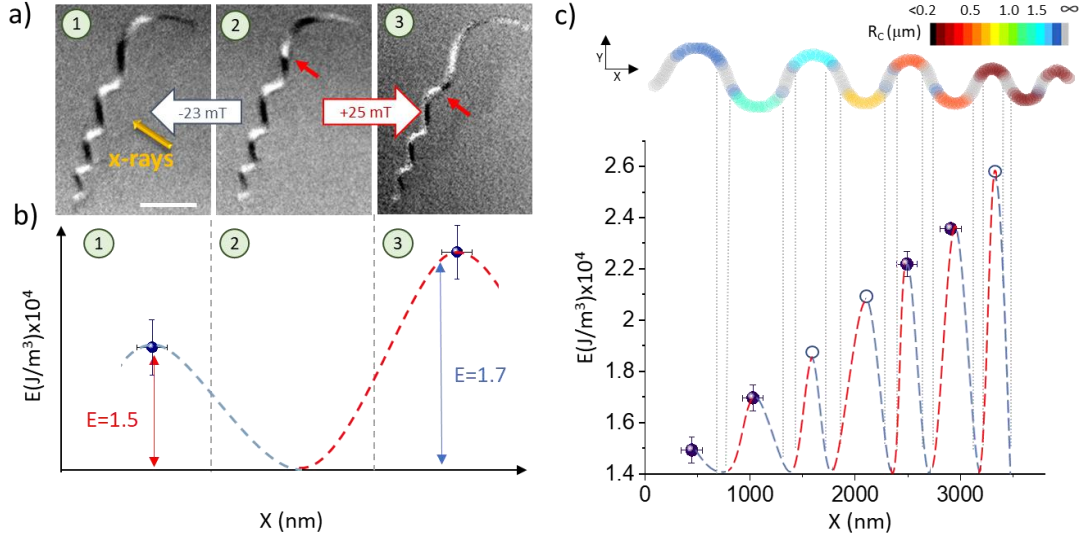

**Figure 5: Non-reciprocal motion of BPDWs arising from symmetric potential barriers and asymmetric potential wells in a curved nanostructure.** a) XMCD images of the structure, identifying a domain wall at a straight pinning point (2), and the propagation to the left (1) and to the right (3) after the application of field pulses of varying magnitude. b) Energy profile of the potential well of the BPDW in the straight region extracted from the depinning fields, showing an asymmetric potential well corresponding to positive (blue dashed line) and negative (red dashed line) domain wall propagation. c) map of the energy landscape of the BPDW as it is propagated through the structure to the right (blue dashed line) and to the left (red dashed lines). Solid circles represent experimentally determined values while empty circles are estimations. See SI for full dataset.

**Conclusions**

In conclusion, we have determined the influence of curvature-induced chirality on the energy landscape of Bloch point singularities in a model three dimensional nanostructure. With advanced 3D nanopatterning, we control the local symmetry-breaking – and therefore the curvature-induced symmetry breaking. In this way, we demonstrate that Bloch point domain walls predominantly exist in straight, achiral regions of the sample, with curved regions of non-zero chirality leading to an increase of their energy.

We exploit this finding by introducing regions of varying curvature to control the energy landscape of the Bloch points, designing well-defined pinning points in the system. These pinning points are potential wells that exist between two energy barriers, whose height is defined by the local magnitude and gradient of curvature. By locally patterning the gradient, we define potential wells that are asymmetric due to neighbouring energy barriers of different heights, that lead to a non-reciprocal motion of the Bloch points. We demonstrate this control of the energy landscape by realising a robust Bloch point shift-register, with tunable depinning fields, and non-reciprocal behaviour.

This insight into the relationship between Bloch point singularities and the symmetry of a system offers an important insight into the energy of these elusive singularities, and a route to control them. Indeed, Bloch point singularities play an important role in the formation of topological textures: with Bloch point-containing chiral bobbers mediating the nucleation of skyrmion tubes from surfaces, and Bloch point-containing magnetic torons predicted to be involved in the formation of magnetic hopfions [47]. As such textures are observed in chiral magnets [48,49], an understanding of the influence of chirality on Bloch point singularities is key to controlling such nucleation processes. Our experimental demonstration that Bloch point singularities are stable in effectively achiral regions, or interfaces between two chiralities, highlights a future direction for the controlled nucleation of complex 3D textures in chiral magnets.

This concept of harnessing patterned curvature to experimentally explore the influence of chirality is not limited to static magnetic configurations: we envisage that this curvature-induced chirality tuning could be used to probe the influence of chirality on magnetisation dynamics such as magnonics, with potential applications beyond magnetism. Indeed, there have been first theoretical predictions that curvature-induced symmetry breaking could lead to tunable chirality in superconductors [50] and superconductor-magnet heterostructures [51], as well as for van der Waal magnets [52]. We envisage that curvature-induced tuning will not only provide key insights into the physics of broader quantum materials, but will provide a route to tune, and control the emergent behaviour for future applications. This control, in combination with more complex three dimensional architectures, will provide new opportunities for high density and interconnected logic devices [54,55].

## Methods

### Fabrication

The 3D cobalt undulating nanostructures were grown using focused electron beam deposition (FEBID) with the Fei Helios G3 Ga-FIB system at the Max Planck Institute for Chemical Physics of Solids in Dresden. The structures were fabricated on top of Si substrates with a 15-nm thick Au layer to improve the conductivity. The structures were designed using FreeCAD software and converted to beam scanning patterns using the f3ast software [31]. The parameters chose for the growth of the structures were an acceleration voltage of 6 kV and a current of 43 pA. For these parameters, the growth times varied from 10 to 20 min. The radius of curvatures along the structures were measured with a Scanning Electron Microscopy (SEM). After deposition, the samples were annealed for 30 min at 250º C in ultra-high vacuum conditions to avoid the deformation of the structures during XMCD-PEEM measurements.

### Shadow-XPEEM

Shadow XMCD-PEEM measurements were performed at CIRCE beamline [34, 35] at Alba synchrotron and at DEMETER beamline at SOLARIS synchrotron. The structures were imaged at the Fe $L_3$-edge measuring images with opposite photon helicities, and subtracted pixel by pixel to determine the in-plane magnetization component along the X-ray direction with nanometer resolution. In plane magnetic fields were applied in situ using the uniaxial sample holder available at CIRCE beamline[i]. The sample holder provides a maximum in-plane magnetic field of 80mT/A. All the XMCD-PEEM images were acquired in remanence after the application of magnetic fields.

### Micromagnetic simulations

Micromagnetic simulations were performed using the finite-element method. In order to investigate the energy of the Bloch-point domain wall depending on the curvature of the sample, a wire with a diameter of 70 nm comprised of two 180 degree arcs with radii 200 nm and 150 nm was considered. This structure was extended by 500 nm long straight sections in both directions to minimize the influence of the magnetic surface charges at the ends of the wire onto the domain-wall energy, see supplementary. An initial transition path for the domain-wall moving from the center of the first arc, to the center of the second arc, was initialized by parametrizing a Bloch-point domain wall at 40 sample points. This initial transition path was numerically evolved employing the string method. In each iteration of the string method, every magnetization configuration in the transition path is driven a certain amount towards the energetic equilibrium using the steepest descent method. The resulting magnetization configurations are then rearranged on the transition path in an equidistant fashion using cubic interpolation..

The energies of the magnetization configurations in the converged transition path are shown in Fig. S4 of the suppplementary. Since the energy of a Bloch-point domain wall is dominated by the exchange contribution of the Bloch point, the numerical results are subject to significant noise caused by the irregular cell sizes in the tetrahegonal finite-element mesh despite the choice of 4 nm as a mesh cell size. This is due to the singular nature of the Bloch point that leads to a systematic underestimation of its exchange energy depending on the size of the containing simulation cell. In order mitigate this noise, the string simulation was repeated with 20 different tetragonalizations of the geometry. The resulting energies were averaged over the mesh realizations.

### Acknowledgements

These experiments were performed at CIRCE beamline at ALBA Synchrotron Light Facility and at DEMETER beamline at National Synchrotron Radiation Centre SOLARIS with the collaboration of ALBA and SOLARIS staff. S. R-G., P. M-F, C. F-G and C.D. acknowledge funding from the Max Planck Society Lise Meitner Excellence Program. P.F-G acknowledges support from the International Max Planck Research School for Chemistry and Physics of Quantum Materials. S. R-G acknowledges support from the Humboldt foundation grant 1223621 and Marie Curie fellowship grant GAP-101061612. C.A. acknowledges funding from FWF through the projects. A. F.-P. Acknowledges funding from the European Community under the Horizon 2020 program. Contract NO 101001290 (3DNANOMAG).

---

i